\documentclass[a4paper,english,a4paper,preprint,showpacs,preprintnumbers,amsmath,amssymb,prb]{revtex4}
\usepackage[latin1]{inputenc}
\usepackage{float}
\usepackage{graphicx}
\usepackage{amssymb}

\makeatletter

\providecommand{\tabularnewline}{\\}

\usepackage{float}

\makeatletter

\usepackage{float}

\makeatletter

\usepackage{dcolumn}

\usepackage{bm}

\makeatother

\makeatother

\usepackage{babel}
\makeatother
\begin{document}

\title{Analysis of resonant inelastic x-ray scattering \\
 at the $K$ edge in NiO }

\author{Manabu Takahashi$^{1}$, Jun-ichi Igarashi$^{2}$, and Takuji Nomura$^{3}$}

\affiliation{$^{1}$Faculty of Engineering, Gunma University, Kiryu, Gunma 376-8515,
Japan\\
 $^{2}$Faculty of Science, Ibaraki University, Mito, Ibaraki 310-8512,
Japan\\
 $^{3}$Synchrotron Radiation Research Center, Japan Atomic Energy
Agency, Hyogo 679-5148, Japan }

\date{\today}

\begin{abstract}
We analyze the resonant inelastic x-ray scattering (RIXS) spectra
at the Ni $K$ edge in an antiferromagnetic insulator NiO by applying
the theory developed by the present authors. It is based on the Keldysh
Green's function formalism, and treats the core-hole potential in
the intermediate state within the Born approximation. We calculate
the single-particle energy bands within the Hartree-Fock approximation
on the basis of the multi-orbital tight-binding model. Using these
energy bands together with the $4p$ density of states from an ab
initio band structure calculation, we calculate the RIXS intensities
as a function of energy loss. By taking account of electron correlation
within the random phase approximation (RPA), we obtain quantitative
agreement with the experimental RIXS spectra, which consist of prominent
two peaks around 5 eV and 8 eV, and the former shows considerable
dispersion while the latter shows no dispersion. We interpret the
peaks as a result of a band-to-band transition augmented by the RPA
correlation. 

\end{abstract}

\pacs{78.70.Ck 71.20.Be 71.28.+d 78.20.Bh}

\maketitle

\section{\label{sect.1}Introduction}

Excitations in solids are fundamental to describe physical properties
such as the response to external perturbations and temperature dependence.
They may be characterized into two types, spin and charge excitations.
For the former, the inelastic neutron scattering is quite powerful
to investigate energy-momentum relations. By contrast, charge excitations
have been investigated by measuring the optical conductivity, but
the momentum transfer is limited to nearly zero. \cite{Uchida91}
The electron energy loss spectroscopy can detect the momentum dependence
of charge excitations, but it suffers from strong multiple scattering
effects.\cite{Wang96} Recently, taking advantage of strong synchrotron
sources, the resonant inelastic x-ray scattering (RIXS) has become
a powerful tool to probe charge excitations in solids. \cite{Kao96,Hill98,Hasan00,Kim02,Inami03,Kim04-1,Suga05}
In transition-metal compounds, $K$-edge resonances are widely used
to observe momentum dependence, because corresponding x-rays have
wavelengths the same order of lattice spacing. The process is described
as a second-order optical process that the $1s$ core electron is
prompted to an empty $4p$ state by absorbing photon, then charge
excitations are created in order to screen the core-hole potential,
and finally the photo-excited $4p$ electron is recombined with the
core hole by emitting photon. In the end, charge excitations are left
with energy and momentum transferred from photon. In cuprates, the
RIXS spectra are found to have clear momentum dependence. \cite{Hasan00,Kim02,Suga05}

For NiO, an pioneering Ni $K$-edge RIXS experiment has been carried
out by Kao et al., in which the spectral peaks were not well resolved
had no clear momentum dependence, probably due to the experimental
resolution. \cite{Kao96} Recently, a new $K$-edge RIXS experiment
has been carried out, at Taiwan beam line in SPring-8.\cite{H.Ishii06}
They have observed the spectra as a function of energy loss by tuning
the incident photon energy at 8351 eV, which corresponds to the Ni
$K$-edge absorption peak. The observed spectra consist of two prominent
peaks at 5 eV and 8 eV, and the 5-eV-peak considerably changes while
the 8-eV-peak does not change with changing momenta. In addition,
extra tiny peaks are found below 4 eV, which are called {}``$d$-$d$
excitation\char`\"{}. In this paper, we analyze the RIXS spectra for
NiO by developing the formalism of Nomura and Igarashi.\cite{Nomura04,Nomura05}
This theory is based on the many-body formalism of Keldysh, and is
regarded as an extension of the resonant Raman theory developed by
Nozi\`eres and Abrahams.\cite{Nozieres74} The core-hole potential
is treated within the Born approximation. Higher-order effects beyond
the Born approximation have been evaluated on the $K$-edge RIXS in
La$_{2}$CuO$_{4}$.\cite{Igarashi06} Although the core-hole potential
is rather strong, the higher-order effects are found to cause only
minor change in the spectral shape.\cite{Igarashi06} In this situation,
the RIXS spectra can be connected to the $3d$-density-density correlation
function in the equilibrium state. We develop the formalism by clarifying
the equivalence of the Keldysh formalism and the conventional Green's
function formalism and by deriving the RIXS formula for possible bound-states
corresponding to the $d$-$d$ excitations. Advantages of the present
formalism are, in contrast to the numerical diagonalization method,
that it is applicable to three-dimensional models consisting of many
orbitals, and that it provides clear physical interpretation.

We construct a detailed multi-orbital tight-binding model including
all $3d$ orbitals as well as the full intra-atomic Coulomb interaction
between $3d$ orbitals. Applying the Hartree-Fock approximation (HFA)
to the model, we obtain the antiferromagnetic (AF) solution with an
appropriate description of the single-particle spectra having a large
energy gap. Unoccupied $3d$ states on a Ni site consist of minority
spin states on the site and have almost the $e_{g}$-character. Note
that the band calculation with the local density approximation (LDA)
fails to reproduce the energy gap of 4 eV. The RIXS spectra are interpreted
as a result of a band-to-band transition to screen the $1s$ core
hole. Therefore the transition occurs through the amplitude of the
$e_{g}$-character. Treating electron correlations by the random phase
approximation (RPA), we find that spectral shape as a function of
energy loss is strongly modified in the continuum spectra. In order
to obtain quantitative agreement with the experiment, we need to take
account of the RPA correlation. Note that the same analysis of the
RIXS in cuprates has been successful to reproduce the experimental
spectra. The success of the analysis for NiO would add another evidence
of the usefulness of the present scheme of analyzing the RIXS spectra.

The present formalism describes the \char`\"{}$d$-$d$\char`\"{}
excitation by the bound state in the density-density correlation function.
We obtain bound states but with extremely small intensities. We will
discuss possible reasons for the discrepancy with the experiment.

The present paper is organized as follows. In Sec. \ref{sect.2},
we study the electronic structure on the basis of the band calculation,
and then introduce a multi-orbital tight-binding model. We calculate
the single-particle Green's function within the HFA in the AFM phase
of NiO. In Sec. \ref{sect.3}, we summarize the formula for the RIXS
spectra, including the discussion of the bound states. In Sec. \ref{sect.4},
we calculate the RIXS spectra by taking account of the RPA correction
in comparison with the experiment. Section \ref{sect.5} is devoted
to the concluding remarks.

\section{\label{sect.2}Electronic structure of Nickel Oxide}

\begin{figure}
\includegraphics[width=7cm,keepaspectratio]{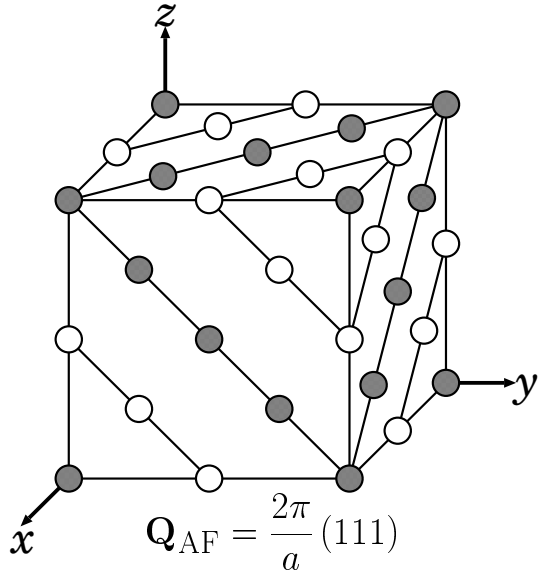}

\caption{\label{fig.crystal} Schematic view of a NiO crystal with type-II
AF order. Only Ni atoms are shown. Wave vector ${\bf Q}_{{\rm AF}}$
characterizes the AF modulation. The direction of magnetic moment
on the Ni site denoted by filled circle is antiparallel to that on
the site denoted by open circle.}
\end{figure}

The crystal structure of NiO is the NaCl-type with the lattice constant
of $a=4.177$ A. Ni atoms form an fcc lattice, as shown in Fig. \ref{fig.crystal}.
Type -II AF order develops below $T_{{\rm N}}=523$ K. The order parameter
is characterized by a wave vector directing to one of four body diagonals
in the fcc lattice.\cite{Roth1958a}

\subsection{Ab initio calculation}

\begin{figure}
\includegraphics[width=8cm]{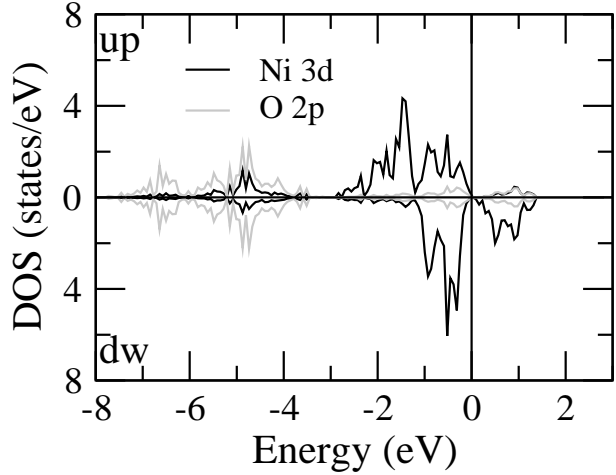}

\caption{\label{fig.lda_dos} Density of states projected onto Ni$3d$ and
O$2p$ states calculated within the LDA. The energy zero is at the
top of the valence band.}
\end{figure}

We calculate the electronic band structure using the muffin-tin KKR
method within the LDA. Although we obtain a stable AF self-consistent
solution, the energy gap is much smaller than the experimental one,
$\sim4.3$ eV.\cite{Sawatzky1984} Figure \ref{fig.lda_dos} shows
the calculated density of states (DOS) projected onto Ni$3d$ and
O$2p$ states. The difference between the calculation and the experiment
in the energy gap may be improved by using the so called LDA$+U$
method, which is a hybrid of the LDA and the HFA for the Coulomb interaction
in the Ni $3d$ orbitals. Since we need to calculate the two-particle
correlation function within the RPA, we introduce a multi-orbital
tight-binding model in place of the ab initio calculation, and apply
the RPA in the following.

\begin{figure}
\includegraphics[width=8cm]{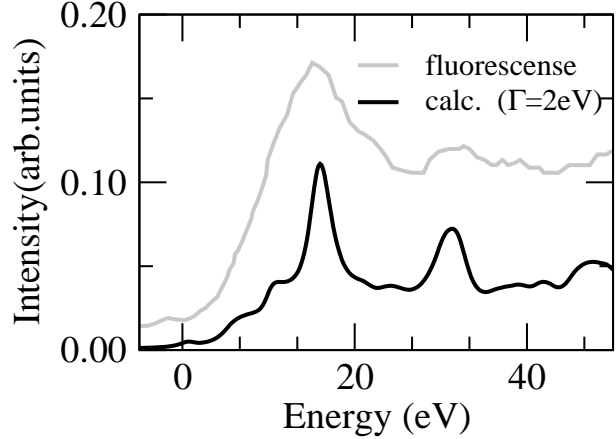}

\caption{\label{fig.lda_4pdos} Density of states projected onto Ni$4p$ states
calculated within the LDA. The experimental curve is taken from Fig.
4 in ref. \onlinecite{Neubeck2001}.}
\end{figure}

In contrast to the failure for the $3d$ states, we expect that the
$4p$ bands are well described within the LDA, since the $4p$ band
has a wide width $\gtrsim20$ eV and thereby are weakly correlated.
We show the $4p$DOS convoluted with the Lorentzian function with
the full width of half maximum (FWHM) 2 eV in Fig. \ref{fig.lda_4pdos}.
The FWHM corresponds to the core-hole life-time width. The calculated
curve agrees fairly well with the experimental one.

\subsection{Multi-orbital tight-binding model}

We introduce a multi-orbital tight-binding model defined by \begin{eqnarray}
H & = & H_{0}+H_{I},\\
H_{0} & = & \sum_{im\sigma}E^{d}n_{im\sigma}^{d}+\sum_{j\ell\sigma}E^{p}n_{j\ell\sigma}^{p}+\sum_{\left\langle i,j\right\rangle }\sum_{\sigma\ell m}\left(t_{im,j\ell}^{dp}d_{im\sigma}^{\dagger}p_{j\ell\sigma}+H.c.\right)\\
 & + & \sum_{\left\langle j,j'\right\rangle }\sum_{\sigma\ell\ell'}\left(t_{j\ell,j'\ell'}^{pp}p_{j\ell\sigma}^{\dagger}p_{j'\ell'\sigma}+H.c.\right)+\sum_{\left\langle i,i'\right\rangle }\sum_{\sigma mm'}\left(t_{im,i'm'}^{dd}d_{im\sigma}^{\dagger}d_{i'm'\sigma}+H.c.\right),\\
H_{I} & = & \frac{1}{2}\sum_{i}\sum_{\nu_{1}\nu_{2}\nu_{3}\nu_{4}}g\left(\nu_{1}\nu_{2};\nu_{3}\nu_{4}\right)d_{i\nu_{1}}^{\dagger}d_{i\nu_{2}}^{\dagger}d_{i\nu_{4}}d_{i\nu_{3}}.\end{eqnarray}
 The part $H_{0}$ represents the kinetic energy, where $d_{im\sigma}$
and $p_{j\ell\sigma}$ denote the annihilation operators of an electron
with spin $\sigma$ in the $3d$ orbit $m$ of Ni site $i$ and the
annihilation operator of an electron with spin $\sigma$ in the $2p$
orbit $\ell$ of the O-site $j$, respectively. Number operators $n_{im\sigma}^{d}$
and $n_{j\ell\sigma}^{p}$ are defined by $n_{im\sigma}^{d}=d_{im\sigma}^{\dagger}d_{im\sigma}$,
$n_{j\ell\sigma}^{p}=p_{j\ell\sigma}^{\dagger}p_{j\ell\sigma}$. The
transfer integrals, $t_{im,j\ell}^{dp}$, $t_{j\ell,j'\ell'}^{pp}$
$t_{im,i'm'}^{dd}$, are evaluated from the Slater-Koster two-center
integrals, $(pd\sigma)$, $(pd\pi)$, $(pp\sigma)$, $(pp\pi)$, $(dd\sigma)$,
$(dd\pi)$, $(dd\delta)$.\cite{Slater54} The $d$-level position
relative to the $p$-levels is given by the charge-transfer energy
$\Delta$ defined by $\Delta=E_{d}-E_{p}+8U$ for the $d^{8}$ configuration.\cite{Mizokawa1996}
Here $U$ is the multiplet-averaged $d$-$d$ Coulomb interaction
given by $U=F^{0}-\left(2/63\right)F^{2}-\left(2/63\right)F^{4}$,
where $F^{0}$, $F^{2}$, and $F^{4}$ are Slater integrals for $3d$
orbitals. The part $H_{I}$ represents the intra-atomic Coulomb interaction
on TM sites. The Coulomb interaction on O sites is neglected. The
interaction matrix element $g\left(\nu_{1}\nu_{2};\nu_{3}\nu_{4}\right)$
is written in terms of $F^{0}$, $F^{2}$, and $F^{4}$ ($\nu$ stands
for $\left(m,\sigma\right)$).

We determine most parameter values from a cluster-model analysis of
photo-emission spectra. \cite{Elp92} The values for $(dd\sigma)$,
$(dd\pi)$, and $(dd\delta)$ can not be determined from the cluster-model
analysis, and therefore we set them close to Mattheiss' LDA estimates.
\cite{Mattheiss72} Among Slater integrals, $F^{2}$ and $F^{4}$
are known to be slightly screened by solid-state effects, so that
we use the values multiplying $0.8$ to atomic values. On the other
hand, $F^{0}$ is known to be considerably screened, so that we regard
the value as an adjustable parameter to get a reasonable band gap.
Table \ref{table.1} lists the parameter values used in the present
calculation. 

\begin{table}

\caption{\label{table.1} Parameter values for the tight-binding model of
NiO in units of eV. }

\begin{tabular}{rrrrr}
\hline 
&
\multicolumn{2}{c}{SK param.}&
\multicolumn{2}{c}{Slater Integral}\tabularnewline
Ni~~~&
$dd\sigma$~~~&
$-0.227$&
~~~~~$F^{0}$&
$5.00$~~~~~~~\tabularnewline
&
$dd\pi$~~~&
$-0.103$&
$F^{2}$&
$10.57$~~~~~~~\tabularnewline
&
$dd\delta$~~~&
$-0.010$&
$F^{4}$&
$7.56$~~~~~~~\tabularnewline
\hline 
Ni-O~~~&
$pd\pi$~~~&
$-1.400$&
\multicolumn{2}{c}{~~~charge transfer energy}\tabularnewline
&
$pd\sigma$~~~&
$0.630$&
$\Delta$&
$4.5$~~~~~~~\tabularnewline
\hline 
O~~~&
$pp\sigma$~~~&
$0.600$&
&
\tabularnewline
&
$pp\pi$~~~&
$-0.150$&
&
\tabularnewline
\hline
\end{tabular}
\end{table}

\subsection{Hartree-Fock Approximation}

For the AFM order shown in Fig.~\ref{fig.crystal}, a unit cell contains
two Ni atoms and two O atoms. Labeling a unit cell by $\eta$, we
introduce the Fourier transform of the annihilation operator $A_{\eta\mu}$
($=d_{im\sigma}$ or $p_{j\ell\sigma}$) in the magnetic Brillouin
zone, \begin{equation}
A_{{\bf k}\mu}=\sqrt{\frac{2}{N}}\sum_{\eta}A_{\eta\mu}{\rm e}^{i{\bf kr}_{\eta}}.\end{equation}
 Here ${\bf r}_{\eta}$ represents a position vector of the unit cell,
and $\eta$ runs over $N/2$ unit cells. Note that a single phase
factor ${\bf kr}_{\eta}$ is applied to all the states in each unit
cell. With these operators, the single-particle Green's function is
introduced in a matrix form, \begin{equation}
[\hat{G}({\bf k},\omega)]_{\mu\mu'}=-i\int\langle T(A_{{\bf k}\mu}(t)A_{{\bf k}\mu'}^{\dagger}(0))\rangle{\rm e}^{i\omega t}{\rm d}t.\end{equation}

In the HFA, we disregard the fluctuation terms in $H_{I}$ and approximate
$H_{I}$ by \begin{equation}
H_{I}^{HF}=\frac{1}{2}\sum_{i}\sum_{\nu_{1}\nu_{2}\nu_{3}\nu_{4}}\Gamma^{(0)}(\nu_{1}\nu_{2};\nu_{3}\nu_{4})\langle d_{i\nu_{2}}^{\dagger}d_{i\nu_{3}}\rangle d_{i\nu_{1}}^{\dagger}d_{i\nu_{4}},\end{equation}
 where $\Gamma^{(0)}$is the antisymmetric vertex function defined
by \begin{equation}
\Gamma^{(0)}(\nu_{1}\nu_{2};\nu_{3}\nu_{4})=g(\nu_{1}\nu_{2};\nu_{3}\nu_{4})-g(\nu_{1}\nu_{2};\nu_{4}\nu_{3}).\end{equation}
 where $\langle A\rangle$ denotes the ground state average of the
operator $A$. With the Hamiltonian $H_{0}+H_{I}^{HF}$, we obtain
the following relation: \begin{equation}
(\omega\hat{I}-\hat{J}({\bf k}))\hat{G}({\bf k},\omega)=\hat{I},\end{equation}
 where $\hat{I}$ is the unit matrix, and $\hat{J}({\bf k})$ is given
by \begin{equation}
[H_{0}+H_{I}^{HF},A_{{\bf k}\mu}]=\sum_{\mu'}J_{\mu\mu'}({\bf k})A_{{\bf k}\mu'}.\end{equation}
 Introducing an unitary matrix $\hat{U}({\bf k})$ to diagonalize
$\hat{J}({\bf k})$, that is, $[\hat{U}({\bf k})^{-1}\hat{J}({\bf k})\hat{U}({\bf k})]_{jj'}=E_{j}({\bf k})\delta_{jj'}$,
we express the Green's function as \begin{equation}
\hat{G}({\bf k},\omega)=\hat{U}({\bf k})\hat{D}({\bf k},\omega)\hat{U}({\bf k})^{-1},\end{equation}
 with \begin{equation}
[\hat{D}({\bf k},\omega)]_{jj'}=\frac{1}{\omega-E_{j}({\bf k})+i\delta{\rm sign}(E_{j}({\bf k})-\mu_{0})}\delta_{jj'},\end{equation}
 where $\mu_{0}$ is the chemical potential. The $\hat{J}({\bf k})$
contains expectation values on the ground state, which should be self-consistently
determined from the relation, \begin{equation}
\langle d_{i\mu}^{\dagger}d_{i\mu'}\rangle=\frac{2}{N}\sum_{{\bf k}}\int[\hat{G}({\bf k},\omega)]_{\xi'\xi}{\rm e}^{i\omega0^{+}}\frac{{\rm d}\omega}{2\pi},\end{equation}
 where $\xi$ and $\xi'$ stand for $\mu$ and $\mu'$, respectively,
at one of two Ni sites in a unit cell. Only the diagonal parts are
non-vanishing, if $3d$ orbitals are specified by $m=xy$, $yz$,
$zx$, $x^{2}-y^{2}$, $3z^{2}-r^{2}$ with $x$, $y$, $z$ referring
to the crystal axes, that is, $\langle d_{i\nu}^{\dagger}d_{i\nu'}\rangle\neq0$
only for $\nu=\nu'$.

\begin{figure}
\includegraphics[width=8cm]{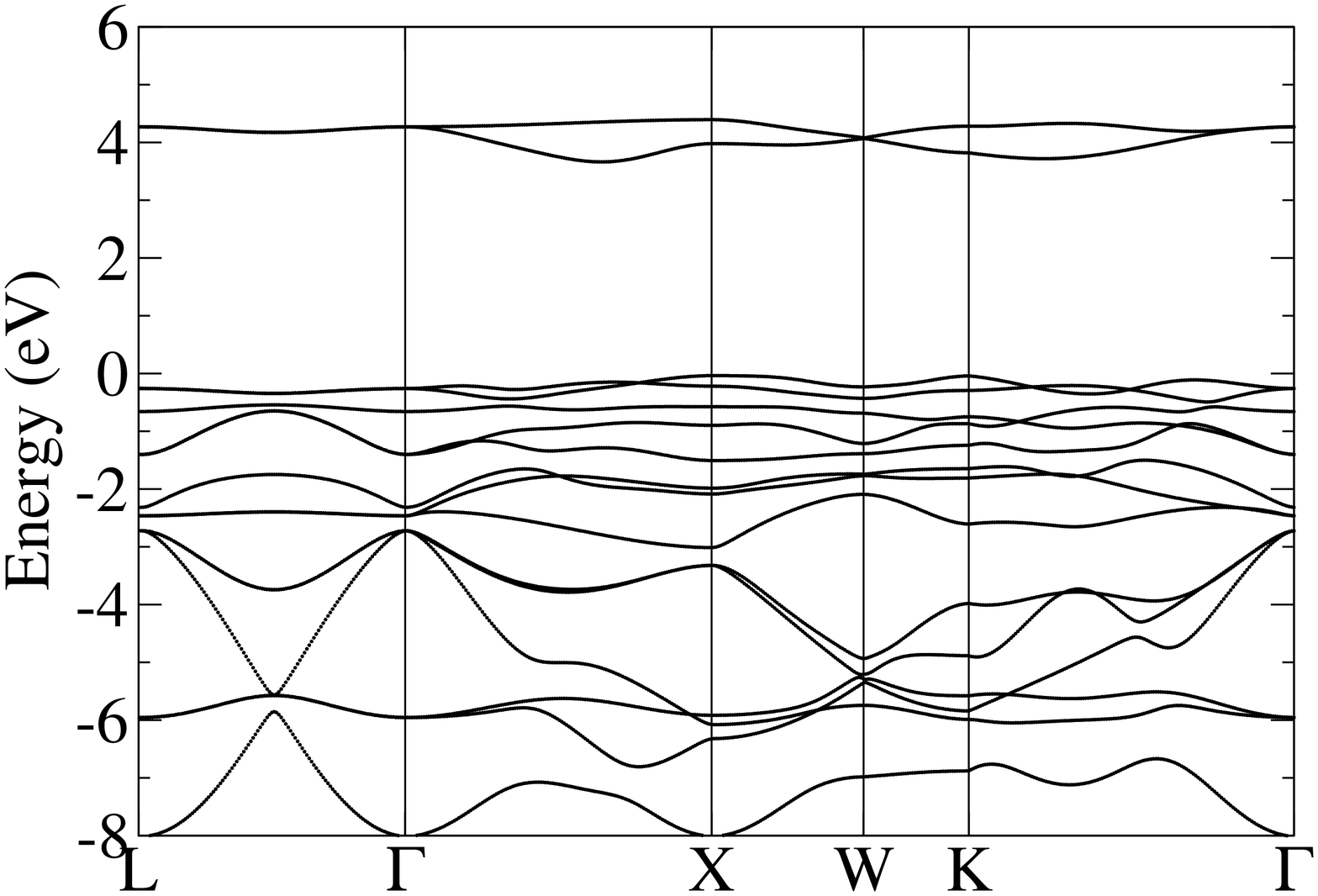}

\caption{\label{fig.hf_disp} Energy band as a function of momentum along
symmetry lines within the HFA. The energy zero is at the top of the
valence band.}
\end{figure}

A stable self-consistent solution exists for the AFM order shown in
Fig.~\ref{fig.crystal}. Figure \ref{fig.hf_disp} shows the energy
band as a function of momentum along symmetry lines. The energy gap
is obtained as $\sim4$ eV, in consistent with the experiment.

\begin{figure}
\includegraphics[width=8cm]{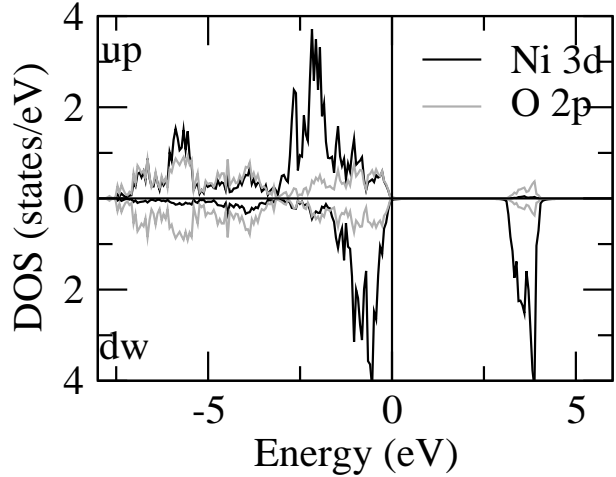}

\caption{\label{fig.hf_dos} Spin-resolved DOS projected onto Ni $3d$ states
and onto O $2p$ states. The energy zero is at the top of the valence
band.}
\end{figure}

\begin{figure}
\includegraphics[width=8cm]{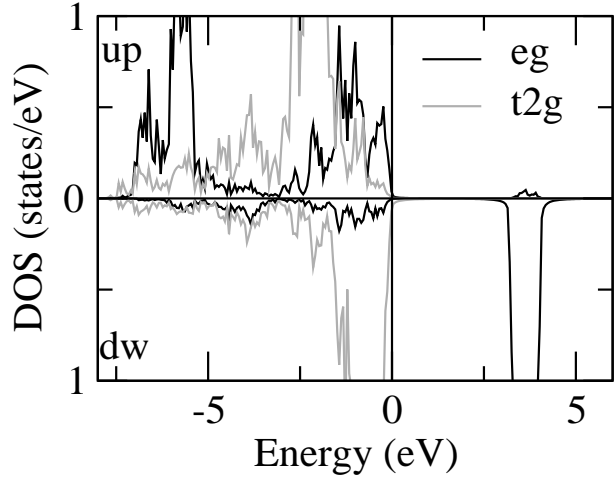}

\caption{\label{fig.hf_t2g_eg_dos} Spin-resolved DOS divided into the $t_{2g}$
and $e_{g}$ characters. The energy zero is at the top of the valence
band.}
\end{figure}

Figures \ref{fig.hf_dos} and \ref{fig.hf_t2g_eg_dos} shows the DOS
projected onto Ni $3d$ and O $2p$ states, and the $3d$DOS divided
into the $e_{g}$ and $t_{2g}$ characters, respectively. The states
around the top of the valence band have relatively large weights of
O $2p$ state, implying that NiO is an insulator of charge transfer
type. Regarding the majority spin states, the $3d$ states are almost
fully occupied; the $e_{g}$ character is concentrated around the
top and bottom of the valence band, while the $t_{2g}$ character
is around the middle of the valence band. Regarding the minority spin
states, the unoccupied states have almost the $e_{g}$ character.
The $t_{2g}$ character dominates around the top of the valence band,
while the $e_{g}$ character are widely distributed with relative
small weight. As become clear later, the electron and the hole created
in the RIXS process have the same spin. Therefore, the pair creation
in the RIXS takes place in the minority spin states with the $e_{g}$
character.

The electron correlation modifies the bands given by the HFA. One
of the most prominent difference is that a {}``satellite\char`\"{}
peak is created around 9 eV below the top of the valence band. In
addition, the $3d$ states are pushed to upper energy position as
a counter effect of the satellite creation. However, these modifications
are limited deep in the low energy part of the valence band, mostly
with the majority spin. Therefore, the electron correlation on the
single-particle excitation may have little influence on the RIXS spectra.

\section{\label{sect.3}Formula for RIXS spectra}

\subsection{general expression}

For the interaction between photon and matter, we consider the dipole
transition at the $K$ edge, where the $1s$-core electron is excited
to the $4p$ band with absorbing photon and the reverse process takes
place. This process may be described by \begin{equation}
H_{x}=w\sum_{{\bf q}\alpha}\frac{1}{\sqrt{2\omega_{{\bf q}}}}\sum_{j\eta\sigma}e_{\eta}^{(\alpha)}p_{j\eta\sigma}^{\prime\dagger}s_{j\sigma}c_{{\bf q}\alpha}{\rm e}^{i{\bf q}\cdot{\bf r}_{j}}+{\rm H.c.},\end{equation}
 where $e_{\eta}^{(\alpha)}$ represents the $\eta$-th component
($\eta=x,y,z$) of two kinds of polarization vectors ($\alpha=1,2$)
of photon. Since the $1s$ state is so localized that the $1s\to4p$
dipole transition matrix element is well approximated as a constant
$w$. Annihilation operators $p_{j\eta\sigma}^{\prime}$ and $s_{j\sigma}$
are for states $4p_{\eta}$ and state $1s$ at Cu site $j$, respectively.
The Hamiltonians for the core electron and for $4p$ electrons are
given by \begin{eqnarray}
H_{1s} & = & \epsilon_{1s}\sum_{j\sigma}s_{j\sigma}^{\dagger}s_{j\sigma},\\
H_{4p} & = & \sum_{{\bf k}\eta\sigma}\epsilon_{4p}^{\eta}({\bf k})p_{{\bf k}\eta\sigma}^{\prime\dagger}p_{{\bf k}\eta\sigma}^{\prime}.\end{eqnarray}
 The photo-created $1s$-core hole induces charge excitations through
the attractive core-hole potential, which may be described by \begin{equation}
H_{1s-3d}=V\sum_{im\sigma\sigma'}d_{im\sigma}^{\dagger}d_{im\sigma}s_{i\sigma'}^{\dagger}s_{i\sigma'}.\end{equation}
 Here $V$ may be $5-10$ eV in NiO.

\begin{figure}
\includegraphics[width=8cm]{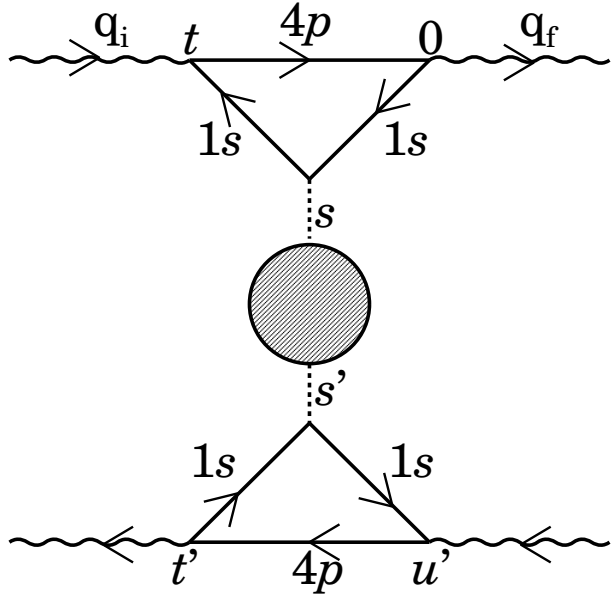}

\caption{\label{fig.diagram1} Diagrams for the RIXS intensity within the
Born approximation to the $1s$ core-hole potential. The dotted lines
represent the core-hole potential $V$. The shaded part represents
the Keldysh-type Green's function, which connects the outward time
leg on the top half and the backward time leg on the bottom half. }
\end{figure}

This process is diagrammatically represented in Fig.~\ref{fig.diagram1},
where the Born approximation is applied to the screening of the core-hole
potential. The shaded part in the figure represents the Keldysh-type
Green's function, \begin{eqnarray}
Y_{\xi',\xi}^{+-}({\bf q},s'-s) & = & \int Y_{\xi',\xi}^{+-}({\bf q},\omega){\rm e}^{-i\omega(s'-s)}\frac{{\rm d}\omega}{2\pi}\\
 & = & \langle\rho_{{\bf q}m'\sigma'}^{\lambda'}(s')\rho_{{\bf q}m\sigma}^{\lambda\dagger}(s)\rangle,\end{eqnarray}
 where superscripts $+$ and $-$ stand for the backward and outward
time legs, respectively.\cite{Landau62} Confining our discussion
to the zero temperature, we take the average over the ground state.
Therefore, $Y^{+-}$ is nothing but a conventional correlation function
in the equilibrium state. The superfix $\lambda$ on the density operator
represents A or B sites of Ni in a unit cell; we have for $\lambda=A$,
\begin{equation}
\rho_{{\bf q}m\sigma}^{A}=\sqrt{\frac{2}{N}}\sum_{{\bf k}}d_{{\bf k+q}m\sigma}^{A\dagger}d_{{\bf k}m\sigma}^{A}.\end{equation}
 The momentum conservation requires the relation ${\bf q}={\bf q}_{i}-{\bf q}_{f}$,
and ${\bf k}$ runs over the magnetic first BZ.

The product of Green's functions for the $4p$ electron and for the
core hole, which are represented by lines with labels {}``4p\char`\"{}
and {}``1s\char`\"{} in the figure, gives simply a factor $\exp[i(\epsilon_{4p}^{\eta}({\bf p})-\epsilon_{1s}-i\Gamma_{1s}-\omega_{i})t]$
on the outward time leg and a factor $\exp[-i(\epsilon_{4p}^{\eta}({\bf p})-\epsilon_{1s}+i\Gamma_{1s}-\omega_{i})(t'-u')]$
on the backward time leg, where $\Gamma_{1s}$ is a lifetime broadening
width of the $1s$ core hole. The Keldysh Green's function carries
the time dependent factor ${\rm e}^{i\omega s}$ to the outward time
leg and ${\rm e}^{-i\omega s'}$ to the backward time leg. Note that
the core-hole potential works only in intervals $[t,0]$ and $[t',u']$.
Integrating the time factor combined to the above product of Green's
functions, with respect to $s$ and $t$ in the region of $t<s<0$,
$-\infty<t<0$, we obtain \begin{eqnarray}
L_{B}^{\eta}(\omega_{i};\omega) & \equiv & V\int_{-\infty}^{0}{\rm d}t\sum_{{\bf p}}\exp[i(\epsilon_{4p}^{\eta}({\bf p})-\epsilon_{1s}-i\Gamma_{1s}-\omega_{i})t]\int_{t}^{0}{\rm d}s\,{\rm e}^{i\omega s}\nonumber \\
 & = & \int\frac{V\rho_{4p}^{\eta}(\epsilon){\rm d}\epsilon}{(\omega_{i}+\epsilon_{1s}+i\Gamma_{1s}-\epsilon)(\omega_{i}-\omega+\epsilon_{1s}+i\Gamma_{1s}-\epsilon)}.\label{eq.born}\end{eqnarray}
 Here the sum over $4p$ states is replaced by the integration of
the $4p$ DOS projected onto the $\eta$ ($=x,y,z$) symmetry, $\rho_{4p}^{\eta}(\epsilon)$.
A similar factor has been derived in third-order perturbation theory
by Abbamonte et al.\cite{Abbamonte99} The integration with respect
to $s'$ and $t'$ in the backward time leg gives the term complex-conjugate
to Eq.~(\ref{eq.born}). The integration with respect to $u'$ give
the energy conservation factor, which guarantees that $\omega$ in
Eq.~(\ref{eq.born}) is the energy loss, $\omega=\omega_{i}-\omega_{f}$.
Finally, combining these relations together, we obtain an expression
of the RIXS intensity for the incident photon with the momentum and
energy $q_{i}=({\bf q}_{i},\omega_{i})$, polarization ${\bf {\rm e}^{(\alpha_{i})}}$,
and the scattered photon with the momentum and energy $q_{f}=({\bf q}_{f},\omega_{f})$,
polarization ${\bf {\rm e}^{(\alpha_{f})}}$: \begin{equation}
W(q_{i},\alpha_{i};q_{f},\alpha_{f})=2\pi\frac{|w|^{4}}{4\omega_{i}\omega_{f}}\sum_{\xi\xi'}Y_{\xi\xi'}^{+-}(q)\left|\sum_{\eta}e_{\eta}^{(\alpha)}L_{B}^{\eta}(\omega_{i};\omega)e_{\eta}^{(\alpha')}\right|^{2}.\label{eq.general}\end{equation}
 Here $q=({\bf q},\omega)$ with ${\bf q}={\bf q}_{i}-{\bf q}_{f}$,
$\omega=\omega_{i}-\omega_{f}$. Suffix $\xi$ and $\xi'$ represent
the $3d$ states distinguishing the A or B Ni sites.

\subsection{The correlation function}

\begin{figure}
\includegraphics[width=0.8\columnwidth]{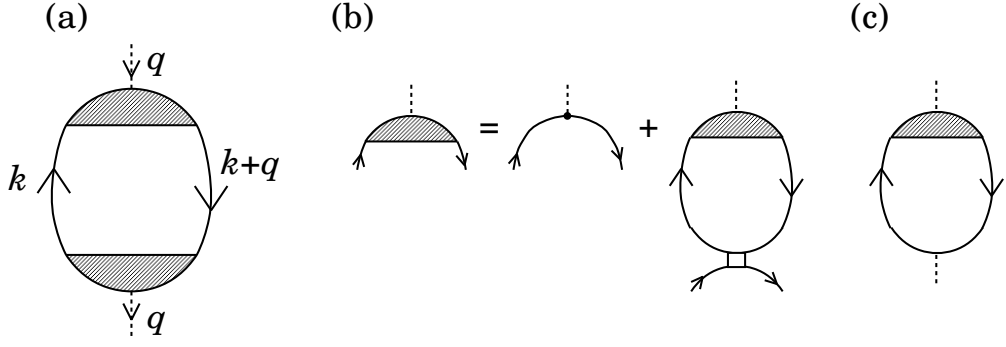}

\caption{\label{fig.diagram2} (a)Diagrams for $Y^{+-}(q)$. Solid lines with
arrows are not the time-ordered Green's functions but the Keldysh-type
ones connecting the outward time leg with the backward time leg. (b)Vertex
function $\Lambda(q)$ within the RPA. Solid lines with arrows are
the conventional time-ordered Green's functions. The square represents
the four-point vertex of the $3d$ Coulomb interaction effective only
at the same Ni sites. (c)Diagrams for the time-ordered $Y^{{\rm T}}(q)$
within the RPA. }
\end{figure}

We consider diagrams where one electron-hole pair remains after the
RIXS process, as shown in Fig.~\ref{fig.diagram2}(a). We
write $Y_{\xi',\xi}^{+-}(q)$ in the form, \begin{equation}
Y_{\xi',\xi}^{+-}(q)=\sum_{\xi_{1}\xi_{2}\xi_{3}\xi_{4}}\Lambda_{\xi_{1}\xi_{2},\xi'}^{*}(q)\Pi_{\xi_{1}\xi_{2},\xi_{3}\xi_{4}}^{+-(0)}(q)\Lambda_{\xi_{3}\xi_{4},\xi}(q),\label{eq.correlation}\end{equation}
 with \begin{eqnarray}
\Pi_{\xi_{1}\xi_{2},\xi_{3}\xi_{4}}^{+-(0)}({\bf q},\omega) & = & 2\pi\frac{2}{N}\sum_{{\bf k}}\sum_{j,j'}\delta(\omega-E_{j'}({\bf k+q})+E_{j}({\bf k}))[1-n_{j'}({\bf k+q})]n_{j}({\bf k})\nonumber \\
 & \times & U_{\xi_{1}j'}({\bf k+q})U_{\xi_{3}j'}^{*}({\bf k+q})U_{\xi_{4}j}({\bf k})U_{\xi_{2}j}^{*}({\bf k}),\label{eq.keldysh1}\end{eqnarray}
 where $j$ and $j'$ denote energy eigenstates. The $\delta$-function
in $\Pi^{+-(0)}(q)$ indicates that the inter band transition from
the valence band to the conduction band gives rise to the RIXS intensity.
Only the weight of $3d$ states in the bands contributes to the intensity.
We evaluate the vertex function in Eq.~(\ref{eq.correlation}) within
the RPA. Collecting the ladder diagrams shown in Fig.~\ref{fig.diagram2}(b),
we obtain \begin{equation}
\Lambda_{\xi_{1}\xi_{2},\xi}(q)=\left[\hat{I}-\hat{\Gamma}\hat{F}^{--}(q)\right]_{\xi_{1}\xi_{2},\xi\xi}^{-1},\end{equation}
 where $\hat{F}^{--}(q)$ is the conventional time-ordered propagator
given by \begin{eqnarray}
[\hat{F}^{--}(q)]_{\xi_{1}\xi_{2},\xi_{3}\xi_{4}} & = & -i\frac{2}{N}\sum_{{\bf k}}\int\frac{{\rm d}k_{0}}{2\pi}G_{\xi_{4}\xi_{2}}({\bf k},k_{0})G_{\xi_{1}\xi_{3}}({\bf k+q},k_{0}+\omega)\nonumber \\
 & = & \frac{2}{N}\sum_{{\bf k}}U_{\xi_{4}j}({\bf k})U_{\xi_{2}j}^{*}({\bf k})U_{\xi_{1}j'}({\bf k+q})U_{\xi_{3}j'}^{*}({\bf k+q})\nonumber \\
 & \times & \left[\frac{n_{j}({\bf k})[1-n_{j'}({\bf k+q})]}{\omega-E_{j'}({\bf k+q})+E_{j}({\bf k})+i\delta}-\frac{n_{j'}({\bf k+q})[1-n_{j}({\bf k})]}{\omega-E_{j'}({\bf k+q})+E_{j}({\bf k})-i\delta}\right].\label{eq.green_pair}\end{eqnarray}
 The four-point vertex is non-zero only for $\xi_{1}$, $\xi_{2}$,
$\xi_{3}$, $\xi_{4}$ belonging to the \emph{same} Ni site, that
is, \begin{equation}
[\hat{\Gamma}]_{\xi_{1}\xi_{2},\xi_{3}\xi_{4}}=\Gamma^{(0)}(\xi_{1}\xi_{4};\xi_{2}\xi_{3}),\end{equation}
 and otherwise it is zero.

As already pointed out, the Keldysh-type Green's function $Y_{\xi',\xi}^{+-}(q)$
is equivalent to the conventional correlation function. Therefore,
it may be more convenient to analyze the function with the help of
the conventional time-ordered Green's function, \begin{equation}
Y_{\xi',\xi}^{{\rm T}}({\bf q},t)=-i\langle T[\rho_{{\bf q}m'\sigma'}^{\lambda'}(t)(\rho_{{\bf q}m\sigma}^{\lambda})^{\dagger}(0)]\rangle,\label{eq.green_time}\end{equation}
 with $T$ being the time-ordering operator. Considering the diagrams
shown in Figs.~\ref{fig.diagram2}(b) and (c), it is expressed
within the RPA as \begin{equation}
Y_{\xi',\xi}^{{\rm T}}(q)=\left\{ \hat{F}^{--}(q)[\hat{I}-\hat{\Gamma}\hat{F}^{--}(q)]^{-1}\right\} _{\xi'\xi',\xi\xi}.\label{eq.time_rpa}\end{equation}
 This expression leads to Eq.~(\ref{eq.correlation}) with the help
of the fluctuation-dissipation theorem (FDT), \begin{equation}
\sum_{\xi'\xi}Y_{\xi'\xi}^{+-}(q)=-i\sum_{\xi'\xi}\left[Y_{\xi'\xi}^{{\rm T}*}(q)-Y_{\xi\xi'}^{{\rm T}}(q)\right]\quad{\rm for}\,\,\omega>0.\label{eq.fdt1}\end{equation}
 To show this fact, we first rewrite Eq.~(\ref{eq.green_pair}) as
\begin{equation}
\hat{F}^{--}(q)=\hat{F}_{1}^{--}(q)+i\hat{F}_{2}^{--}(q),\end{equation}
 where $\hat{F}_{1}^{--}(q)$ and $\hat{F}_{2}^{--}(q)$ are Hermitian
matrices. Then, using the Hermitian property, we transform $Y_{\xi'\xi}^{{\rm T}}(q)^{*}$
as \begin{eqnarray}
Y_{\xi'\xi}^{{\rm T}}(q)^{*} & = & \bigl\{[I-(F_{1}^{--}(q)-iF_{2}^{--}(q))\Gamma]^{-1}[F_{1}^{--}(q)-iF_{2}^{--}(q)]\nonumber \\
 & \times & [I-\Gamma(F_{1}^{--}(q)+iF_{2}^{--}(q))][I-\Gamma(F_{1}^{--}(q)+iF_{2}^{--}(q))]^{-1}\bigl\}_{\xi\xi'}.\end{eqnarray}
 Combining the similar expression for $Y_{\xi\xi'}^{{\rm T}}(q)$,
we have \begin{equation}
Y_{\xi'\xi}^{{\rm T}*}(q)-Y_{\xi\xi'}^{{\rm T}}(q)=\bigl\{[I-(F_{1}^{--}(q)-iF_{2}^{--}(q))\Gamma]^{-1}(-2i)[\hat{F}_{2}^{--}(q)][I-\Gamma(F_{1}^{--}(q)+iF_{2}^{--}(q))]^{-1}\bigl\}_{\xi\xi'}.\label{eq.fdt2}\end{equation}
 Since $-2F_{2}^{--}(q)$ is equivalent to $\Pi^{(0)}(q)$ for $\omega>0$,
the RXS intensity given by Eq.~(\ref{eq.fdt2}) through Eq.~(\ref{eq.fdt1})
is equivalent to the one given by Eq.~(\ref{eq.correlation}).

Equation (\ref{eq.time_rpa}) has sometimes poles for some frequencies
below the energy continuum of an electron-hole pair. These bound states
may be called as {}``$d$-$d$ excitations\char`\"{}, and give rise
to extra RIXS peaks. To analyze the problem, we rewrite Eq.~(\ref{eq.time_rpa})
as \begin{equation}
Y_{\xi',\xi}^{{\rm T}}(q)=\left[\hat{F}^{--}(q)^{-1}-\hat{\Gamma}\right]_{\xi'\xi',\xi\xi}^{-1},\label{eq.bound}\end{equation}
 where ${\hat{F}}^{--}(q)^{-1}$ becomes an Hermitian matrix for $\omega$
below the energy continuum. Therefore, ${\hat{F}}^{--}(q)^{-1}-\Gamma$
can be diagonalized by a unitary matrix. Let the diagonalized matrix
have a zero component at $\omega=\omega_{B}({\bf q})$ with the corresponding
eigenvector $B_{\xi_{1}\xi_{2}}({\bf q})$. Then, $Y_{\xi'\xi}^{{\rm T}}(q)$
can be expanded around $\omega\sim\omega_{B}({\bf q})$ as \begin{equation}
Y_{\xi'\xi}^{{\rm T}}(q)=\frac{C_{\xi'\xi}({\bf q})}{\omega-\omega_{B}({\bf q})},\end{equation}
 with \begin{equation}
C_{\xi'\xi}({\bf q})=\frac{B_{\xi'\xi'}({\bf q})B_{\xi\xi}^{*}({\bf q})}{\sum_{\xi_{1}\xi_{2}\xi_{3}\xi_{4}}B_{\xi_{1}\xi_{2}}^{*}({\bf q})\frac{\partial}{\partial\omega}[\hat{F}^{--}(q)^{-1}]_{\xi_{1}\xi_{2},\xi_{3}\xi_{4}}B_{\xi_{3}\xi_{4}}({\bf q})}.\end{equation}
 Substituting the right hand side of Eq.~(\ref{eq.fdt1}) by this
equation, we obtain \begin{equation}
\sum_{\xi'\xi}Y_{\xi'\xi}^{+-}(q)=2\pi\sum_{\xi'\xi}C_{\xi'\xi}(q)\delta(\omega-\omega_{B}({\bf q})).\end{equation}
 This relation is to be inserted into Eq.~(\ref{eq.general}) for
evaluating the RIXS spectra.

\section{\label{sect.4}Calculated Results}

\begin{figure}[H]
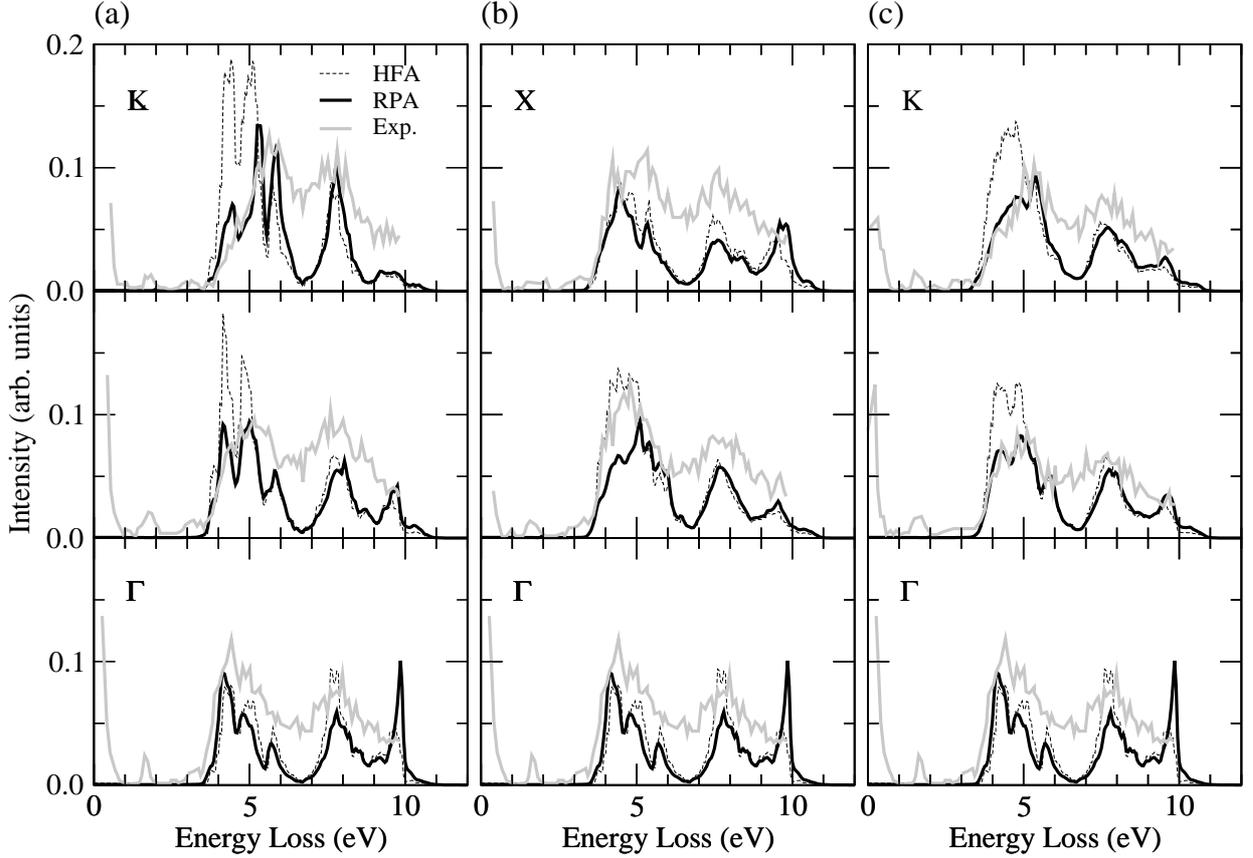

\includegraphics[scale=0.65]{spec005_4_rix_rpa_exp_GL_bw}\includegraphics[scale=0.65]{spec005_4_rix_rpa_exp_GX_bw}\includegraphics[scale=0.65]{spec005_4_rix_rpa_exp_GK_bw}

\caption{\label{fig.spec1} RIXS spectra as a function of energy loss $\omega=\omega_{i}-\omega_{f}$,
in comparison with the experiment by Ishii et al.\cite{H.Ishii06}
Momentum transfer is along the symmetry lines: (a) $\Gamma$- $\frac{2\pi}{a}(0.25,0.25,0.25)$-$L$,
(b) $\Gamma$-$\frac{2\pi}{a}(0.3,0,0)$-$X$, and (c) $\Gamma$-$\frac{2\pi}{a}(0.25,0.25,0)$-$K$.
The incident photon energy is tuned to excite the $1s$ electron to
the peak of the $4p$-DOS. The broken and solid lines are the results
of the HFA and RPA, respectively. The dotted lines represent the experimental
data.\cite{H.Ishii06}}
\end{figure}

We calculate the RIXS intensity from Eq.~(\ref{eq.general}). The
$\delta$-function in $Y^{+-}(q)$ is replaced by the Lorentzian function
with the FWHM $=0.2$ eV in order to take account of the instrumental
resolution. Corresponding to the experimental situation that the polarization
of photon is not specified, we use the realistic $4p$-DOS in evaluating
$L_{B}^{\eta}(\omega_{i};\omega)$ (Eq.~(\ref{eq.born})) and tune
the incident photon energy on the top of the $4p$-DOS. Figure \ref{fig.spec1}
shows the calculated spectra as a function of energy loss in comparison
with the experiment by Ishii et al.\cite{H.Ishii06} The momentum
transfer is chosen along symmetry axes in the Brillouin zone.

The spectra arise from the excitation from the valence states with
the $e_{g}$ character to the conduction states with also the $e_{g}$
character. Since the unoccupied states are almost concentrated in
the minority spin states (see Fig.~\ref{fig.hf_t2g_eg_dos}), the
electron-hole pair creation occurs in the minority spin channel. They
form a continuum spectrum. Several peaks are already present within
the HFA. Some peaks are suppressed and some are enhanced by the RPA.
At the $\Gamma$ point, peaks are found around 4-6 eV and around 8
eV, in good agreement with the experiment. In addition, an extra peak
is found around the upper edge of the continuum, $\sim10$ eV, which
is not confirmed by the experiment (the observation is limited below
10 eV). With momenta deviating from the $\Gamma$ point, intensities
around 5-6 eV is particularly enhanced in the entire region of 4-6
eV. Thereby the spectral shape looks like the peak position is moving
with changing momenta. On the other hand, the peak around 8 eV does
not change its position with changing momenta. These characteristics
are the same along the three symmetry lines. The spectral shapes show
excellent agreement with the experiment. \cite{H.Ishii06} Note that
the sharp peak around 10 eV is strongly suppressed with deviating
momenta from the $\Gamma$ point.

\begin{figure}[H]
 \includegraphics[clip,scale=0.65]{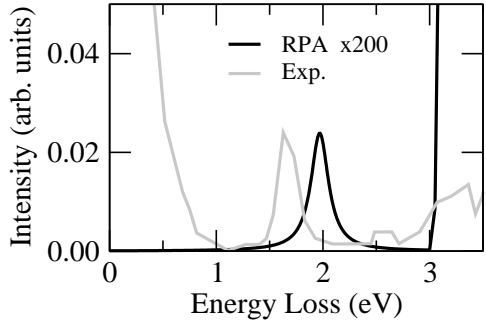}

\caption{\label{fig.spec_bound} RIXS intensity below the lower edge of the
continuum at $\Gamma$ point in comparison with the experiment. \cite{H.Ishii06}
The calculated intensity is enlarged by multiplying 200 to make the
spectra visible.}
\end{figure}

Below the lower edge of the continuum spectra, we obtain a bound state
at 2 eV, as shown in Fig.~\ref{fig.spec_bound}. This may be compared
with the experimental peak at 1.7 eV, but the calculated intensity
is about two order of magnitude smaller. At present, we do not know
the real origin for this discrepancy.

\section{\label{sect.5}Concluding Remarks}

We have analyzed the RIXS spectra in NiO by developing the formalism
of Nomura and Igarashi. This is based on the Keldysh Green's function
formalism and relates the RIXS spectra to the density-density correlation
function under the Born approximation to the core-hole potential.
The use of the Born approximation had been justified for the RIXS
in La$_{2}$CuO$_{4}$ by evaluating the multiple scattering effects.\cite{Igarashi06}
Since NiO has a larger energy gap than La$_{2}$CuO$_{4}$, we expect
that the Born approximation is better justified in NiO. We have introduced
the tight-binding model including all the Ni $3d$ and O $2p$ orbitals
as well as the full Coulomb interaction between $3d$ orbitals, and
have calculated the single-particle excitation within the HFA. The
RIXS spectra are generated by the band-to-band transition to screen
the $1s$ core hole. We have calculated the RIXS spectra using the
single-particle band thus evaluated together with the $4p$ DOS by
the LDA. The electron correlation is treated within the RPA. We have
obtained several peaks in the continuum spectra. Two peaks are found
prominent around 5 eV and 8 eV, and the one around 5 eV shifts to
the higher energy position while another around 8 eV does not move
with momenta changing to the zone boundary. The calculated spectral
shape show quantitative agreement with the experiment. The excellent
agreement with the experiment suggests that the present scheme is
useful to analyze the RIXS spectra. As regards the \char`\"{}$d$-$d$\char`\"{}
excitation, we have obtained a bound state below the lower edge of
the continuum spectrum. However, the intensity is about two order
of magnitude smaller than the experimental one. The presence of crystal
distortion might enhance the intensity of \char`\"{}$d$-$d$\char`\"{}
excitation. At any rate, the real reason for this discrepancy is not
known.

\begin{acknowledgments}
We would like to thank Dr. H. Ishii for providing us with the experimental
data prior to the publication and for valuable discussions. This work
was partially supported by a Grant-in-Aid for Scientific Research
from the Ministry of Education, Culture, Sports, Science, and Technology,
Japan. 
\end{acknowledgments}
\bibliographystyle{apsrev} \bibliographystyle{apsrev}
\bibliography{paper1}

\end{document}